\documentstyle[12pt]{article}
\newcommand{\nc}{\newcommand}
\nc{\lsp}{\;\;\;\;\;\;\;\;}
\nc{\pbarn}{\;\hbox {pb}}
\nc{\lumun}{\;{\hbox {pb}^{-1}}{\hbox {yr}^{-1}}}
\nc{\hc}{\hbox {h.c.}}
\nc{\re}{\hbox {Re}}
\nc{\im}{\hbox {Im}}
\nc{\mev}{\hbox {MeV}}
\nc{\gev}{\;\hbox {GeV}}
\nc{\tev}{\;\hbox {TeV}}
\nc{\etal}{\hbox{et al.}}
\def\gesim{\,{\raise-3pt\hbox{$\sim$}}\!\!\!\!\!{\raise2pt\hbox{$>$}}\,}
\def\lesim{\,{\raise-3pt\hbox{$\sim$}}\!\!\!\!\!{\raise2pt\hbox{$<$}}\,}
\nc{\prd}[3]{{ \it Phys.~Rev.}~{\bf D{#1} (#2) #3}}
\nc{\prl}[3]{{ \it Phys.~Rev.~Lett.}~{\bf {#1} (#2) #3}}
\nc{\plb}[3]{{ \it Phys.~Lett.}~{\bf {#1B} (#2) #3}}
\nc{\npb}[3]{{ \it Nucl.~Phys.}~{\bf B{#1} (#2) #3}}
\nc{\ptp}[3]{{ \it Prog.~Theor.~Phys.}~{\bf {#1} (#2) #3}}
\nc{\zfp}[3]{{ \it Z.~Phys.}~{\bf C{#1} (#2) #3}}
\nc{\mpla}[3]{{ \it Mod.~Phys.~Lett.}~{\bf {#1A} (#2) #3}}
\nc{\rmp}[3]{{ \it Rev.~Mod.~Phys.}~{\bf {#1} (#2) #3}}
\nc{\ijmpa}[3]{{\it Int.~J.~of~Mod.~Phys.}~{\bf {A#1} (#2) #3}}
\nc{\beq}{\begin{equation}}
\nc{\eeq}{\end{equation}}
\nc{\bea}{\begin{eqnarray}}
\nc{\eea}{\end{eqnarray}}
\nc{\baa}{\begin{array}}
\nc{\eaa}{\end{array}}
\nc{\non}{\nonumber}
\nc{\ra} {\rightarrow}
\nc{\ttbar}{\bar{t}t}
\nc{\bbbar}{\bar{b}b}
\nc{\tanb} {\tan \beta}
\nc{\twbdec} {t\rightarrow W^+ b}
\nc{\tbwbdec} {\bar{t} \rightarrow W^- \bar{b}}
\nc{\epem} {e^+e^-}
\nc{\eett} {\epem \ra \ttbar}
\nc{\wpwm} {W^+W^-}
\nc{\tbar} {\bar{t}}
\nc{\bbar} {\bar{b}}
\nc{\wpp} {W^+}
\nc{\mt}{m_t}
\nc{\mts}{m_t^2}
\nc{\mw} {m_W}
\nc{\mws} {m_W^2}
\nc{\mz} {m_Z}
\nc{\mzs} {m_Z^2}
\nc{\ttbardec}{\ttbar \ra W^+W^-\bbbar}
\nc{\wwbb}{W^+W^-\bbbar}
\nc{\bt}{\beta_t}
\nc{\bts}{\beta_t^2}
\nc{\bz}{\beta_Z}
\nc{\bzs}{\beta_Z^2}
\nc{\bw}{\beta_W}
\nc{\bws}{\beta_W^2}
\nc{\nsd} {N_{SD}}
\nc{\ntt} {N_{tt}}
\nc{\lcal}{{\cal{L}}}
\nc{\ocal}{{\cal{O}}}
\nc{\leff}{\lcal_{eff}}
\nc{\sm}{SM}
\nc{\ct}{\cos\theta}
\nc{\cw}{\cos\theta_W}
\nc{\cts}{\cos^2\theta}
\nc{\st}{\sin\theta}
\nc{\sw}{\sin\theta_W}
\nc{\sws}{\sin^2\theta_W}
\nc{\ltv}{\Lambda_{TeV}}
\nc{\lam}{\Lambda}
\nc{\lams}{\lam^2}
\nc{\sig}{\sigma_{tot}}
\nc{\fba}{{{\cal A}}_{FB}}
\nc{\dsig}{\frac{d\sigma}{d\ct}}
\nc{\nsdsig}{N_{SD}^\sigma}
\nc{\nsdasy}{N_{SD}^{FB}}
\nc{\epsl}{\varepsilon_L}

\begin{document}
\begin{flushright}
%Version: \today\hfill
\end{flushright}
\vspace*{2cm}
\begin{center}
{\large  {\bf Four-Fermi Effective Operators at {\boldmath $\eett$}}~\footnote
{Supported in part by the Committee for
Scientific Research, Poland.}\\
\vspace*{2cm}
{\bf Bohdan Grz\c{a}dkowski}\footnote{E-mail:{\tt bohdang@fuw.edu.pl.}}\\
\vspace{1cm}
        Institute for Theoretical Physics\\
        Warsaw University\\ \vspace{1mm}
        Ho\.{z}a 69, PL-00-681 Warsaw, Poland}\\
\vspace*{3cm}

{\bf Abstract}
\end{center}
\vspace{5mm}
The process of top quark pair production at Next Linear Collider (NLC)
has been considered adopting an effective Lagrangian approach
and including
all operators of dim~6 which can be tree-level-generated within unknown
underlying
theory. All contributing helicity amplitudes are presented. It has been found
that four-fermion operators can provide the leading non-standard contribution
to the total cross section.
Expected statistical significance of the non-standard
signal for the total cross section and forward-backward asymmetry have been
calculated taking into account existing experimental constraints.
It has been shown that adopting realistic luminosity of NLC
and  conservative efficiency for the top-quark pair detection,
the total cross section may be sensitive to non-standard physics of an
energy scale around $\Lambda=5\tev$.

\vspace*{2.0cm}
\begin{flushright}
\parbox{2.in}
{IFT 17/95\\
arch-ive/9511279 \\
November 1995}
\end{flushright}
\setcounter{page}{0}
\thispagestyle{empty}
\newpage

\section{Introduction}

Linear high-energy $\epem$ collider can prove to be
very useful laboratory
to study physics of the top quark.
In spite of spectacular successes of experimental high-energy
physics (e.g. precision tests of the Standard Model)
the interactions of the recently discovered~\cite{CDF} top quark are
still unknown. There is no evidence that the top quark interactions
obey the scheme provided by the Standard Model (SM) of electroweak
interactions.
The aim of this letter is to look for non-standard physics effects
in the process $\eett$.
We will show that the present knowledge of the electroweak physics
allows for large beyond the SM corrections to the top
quark production at NLC.
There exist already a large literature~\cite{top}
devoted to non-standard effects
in the top-quark physics, however authors restricts their research to
corrections to the $\ttbar Z$ and/or $\ttbar \gamma$.
There is however no reason to neglect four-Fermi operators which may also
influence the top quark production process $\eett$ at NLC. This is a subject
of the presented research.

We shell follow here
a model independent approach where all possible
non-standard effects are parameterized by means of an effective
Lagrangian~\cite{leffref,gi}. This formalism is model and process independent
and thus
provides an unprejudiced analysis of the data.
We will parameterize all non-standard effects using
the coefficients of a set of effective operators (which respect the
symmetries of the \sm). These operators are
chosen so that there are no {\it a-priori}
reasons to suppose that the said coefficients are suppressed.

The effective Lagrangian approach requires a choice of the low energy
particle content. In this paper we will assume that the \sm\ correctly
describes all such excitations. Thus we
imagine that there is a scale $ \Lambda $, independent of the Fermi
scale, at which the new physics becomes apparent. Since the \sm\ is
renormalizable and the new physics is assumed to be heavy due to a large
dimensional parameter $ \Lambda $, the decoupling
theorem~\cite{Appelquist} is applicable and requires
that all new physics effects be suppressed by inverse powers of $
\Lambda $. All such effects are expressed in terms of a series of local
{\it gauge invariant}
operators of canonical dimension $ > 4 $; the catalogue of such
operators up to dimension 6 is given in Ref.~\cite{bw} (there are no
dimension 5 operators respecting the global and local symmetries of the
\sm).

For the situation we are considering it is natural to assume that the
underlying theory is weakly coupled.
{\it Thus the relevant property of a given
dimension 6 operator is whether it can be generated at tree level by
the underlying physics.} The coefficient of such operators are expected
to be $ O ( 1 ) $; in contrast, the coefficients of loop-generated
operators will contain a suppression~\footnote{If there is a
large number of loop graphs this suppression
factor can be reduced.} factor
$ \sim 1 / 16 \pi^2 $.
The determination of those operators which are tree-level-generated is
given in Ref.~\cite{Arzt.et.al.}.

Here, we shell restrict ourself to effects produced by those operators
which can be tree-level-generated and therefore coefficients of such
operators are not {\it a-priori} suppressed.
In this
respect the present analysis differs from others appearing in the
literature~\cite{others} which concentrate on one-loop-generated
operators related to the
vector-boson self interactions.
(drawbacks of those analysis have been emphasized in Ref.~\cite{gi}).

The strategy which we follow in this paper is to develop the effects of
the tree-level-generated operators contributing to $\ttbar$ production
at $\epem$ collisions. We will consider the constraints implied by current
high-precision data and predict the sensitivity to new effects
at proposed version of NLC.

\section{The Effective Lagrangian}

Hereafter we will adopt a notation from the B\"uchmuller and Wyler classical
paper, see Ref.~\cite{bw}.
The tree-level-generated operators
which will directly contribute to $\eett$ are the following
\beq
\baa[c]{ll}
\ocal^{(1)}_{lq}=\left(\bar{l} \gamma_\mu l \right) \left(\bar{q} \gamma^\mu
q \right) &
\ocal^{(3)}_{lq}=\left(\bar{l} \gamma_\mu \tau^I l \right)
\left(\bar{q} \gamma^\mu \tau^I q \right)  \\
\ocal_{eu}=\left(\bar e \gamma_\mu e \right) \left(\bar u \gamma^\mu u
\right) &
\ocal_{lu}=\left(\bar l u \right) \left( \bar u l \right) \\
\ocal_{qe}=\left(\bar q e \right) \left( \bar e q \right) &
\ocal_{qde}=\left(\bar l e \right) \left( \bar d q \right)\\
\ocal_{lq}=\left(\bar l e \right) \epsilon \left( \bar q u \right) &
\ocal_{lq'}=\left(\bar l u \right) \epsilon \left( \bar q e \right).
\eaa
\label{fermi_oper}
\eeq

The tree-level-generated
operators which modify $\ttbar Z$ and $\ttbar \gamma$ vertices are
\beq
\baa[c]{ll}
\ocal^{(1)}_{\phi q} = i \left( \phi^\dagger D_\mu \phi \right)
                           \left( \bar q \gamma^\mu q \right) &
\ocal^{(3)}_{\phi l} = i \left( \phi^\dagger \tau^I D_\mu \phi \right)
                             \left( \bar q \tau^I \gamma^\mu q \right) \non \\
\ocal_{\phi u} = i \left( \phi^\dagger D_\mu \phi \right)
                           \left( \bar u \gamma^\mu u \right). &
\eaa
\label{ver_oper}
\eeq
The above operators would effect the \sm\
$\ttbar Z$ and $\ttbar \gamma$ vertices and modify the amplitude
for $\eett$ by the s-channel Z and $\gamma$ exchange.

Since we are restricting ourself to tree-level-generated operators,
only vector and axial form-factors for $\ttbar Z$ and $\ttbar \gamma$
vertices could receive any corrections. Therefore we may parameterize
those vertices as :
\beq
\Gamma^i_\mu=\frac{g}{2}\bar t \gamma_\mu (A^i-B^i\gamma_5) t,
\eeq
where $i=\gamma,Z$.
$A^i$ and $B^i$ can be written as a sum of the SM ($A^i_{SM}$, $B^i_{SM}$)
and non-standard ($\delta A^i$, $\delta B^i$) contributions:
\beq
\baa[c]{ll}
%% FOLLOWING LINE CANNOT BE BROKEN BEFORE 80 CHAR
A^\gamma_{SM}=\frac{4}{3}\sw&A^Z_{SM}=\frac{1}{\cw}\left(\frac{1}{2}-\frac{4}{3}\sws\right)\\
B^\gamma_{SM}=0&B^Z_{SM}=\frac{1}{2\cw}\\
\delta A^\gamma=0&\delta A^Z=\frac{1}{2\cw}\left(-\alpha^{(1)}_{\phi q}+
\alpha^{(3)}_{\phi q}-\alpha_{\phi u}\right)\frac{v^2}{\lams}\\
\delta B^\gamma=0&
\delta B^Z=\frac{1}{2\cw}\left(-\alpha^{(1)}_{\phi q}+
\alpha^{(3)}_{\phi q}+\alpha_{\phi u}\right)\frac{v^2}{\lams},
\eaa
\eeq
where $\theta_W$ is the Weinberg angle and $v=246\gev$.

The input parameters we are using are $G_F$, $M_Z$ and $\alpha_{QED}$.
There are tree-level-generated
operators which enter our calculations only through
corrections to the input parameters:
\beq
\baa[c]{ll}
\ocal^{(1)}_\phi = \left( \phi^\dagger \phi \right) \left[ \left( D_\mu
                   \phi \right)^\dagger  D^\mu \phi \right] &
\ocal^{(3)}_\phi = \left( \phi^\dagger D_\mu \phi \right) \left[ \left(
                    D_\mu \phi \right)^\dagger  \phi \right] \\
\ocal^{(3)}_{ll} =  \left( \bar l \tau^I \gamma_\mu l \right)
                    \left( \bar l \tau^I \gamma^\mu l \right) &
\ocal^{(3)}_{\phi l} = i \left( \phi^\dagger \tau^I D_\mu \phi \right)
                             \left( \bar l \tau^I \gamma^\mu l \right).
\eaa
\label{input_oper}
\eeq
Explicit corrections to $G_F$, $M_Z$ and $\alpha_{QED}$ can be obtained
from Ref.~\cite{bw}.

The complete list
of tree-level-generated operators may be found in Ref.~\cite{Arzt.et.al.}.
{\it The effects of those operators present the
widest window into physics beyond the \sm.}

Given the above list the Lagrangian which we will use in the following
calculations is
\beq
\lcal = \lcal^{SM} + \frac{1}{ \Lambda^2} \sum_i\left\{ \alpha_i \ocal_i +
          \hbox{ h.c.} \right\}
\label{lagrangian}
\eeq

It would be more useful
to rewrite (after some necessary Fiertz transformation)
the above four-Fermi operators (\ref{fermi_oper}) in the
following way:
\bea
{\cal{L}}^{4-Fermi}&=&\sum_{i,j=L,R}\left[
S_{ij}\left(\bar{e} P_i e\right) \left(\bar{t} P_j t\right)+
V_{ij}\left(\bar{e}\gamma_\mu P_i e\right)\left(\bar{t}\gamma^\mu P_j t\right)+
\right. \non \\
& &\left. T_{ij}\left(\bar{e} \;\;\frac{\sigma_{\mu \nu}}{\sqrt{2}} \;P_i
e\right)
\left(\bar{t} \;\frac{\sigma^{\mu \nu}}{\sqrt{2}}\; P_j t\right)\right],
\label{fermi_symm}
\eea
where $P_{L,R}=1/2(1\mp\gamma_5)$.
The following constraints must be satisfied by
the coefficients:
\beq
S_{LL}=S_{RR}^\star \lsp V_{ij}=V_{ij}^\star \lsp T_{LL}=T_{RR}^\star \lsp
T_{LR}=T_{RL}=0
\label{constr}
\eeq
{}From the gauge invariance of the Lagrangian it follows that
$S_{LR}=S_{RL}=0$.
All $S_{ij}$, $V_{ij}$ and $T_{ij}$ could be expressed in terms of the initial
$\alpha$'s from the four-Fermi part of the Lagrangian (\ref{lagrangian}).

Apart of the common factor $2iE$, where $E$ is the beam energy,
with $s=4E^2$, $k= E\sqrt{1-4\mts/s}$ and
$\theta$ defined as an angle between outgoing top and incoming electron,
helicity amplitudes for $\eett$
emerging from scalar and tensor type operators read as follows:
\beq
\baa[c]{lll}
(----)&=&(E-k)(S_{LL}+2T_{LL}\ct)-S_{LR}(E+k)\\
(---+)&=&+2T_{LL}\mt\st\\
(--+-)&=&+2T_{LL}\mt\st\\
(--++)&=&(E+k)(S_{LL}-2T_{LL}\ct)-S_{LR}(E-k)\\
(++--)&=&(E+k)(S_{RR}-2T_{RR}\ct)-S_{RL}(E-k)\\
(++-+)&=&-2T_{RR}\mt\st\\
(+++-)&=&-2T_{RR}\mt\st\\
(++++)&=&(E-k)(S_{RR}+2T_{RR}\ct)-S_{RL}(E+k),\\
\eaa
\label{st_amp}
\eeq
where helicities of $e^-$, $e^+$, $t$ and $\bar t$ are indicated in the
parenthesis. For the top quark we use $\mt=174 \gev$.
Vector type operators produce the following amplitudes:
\beq
\baa[c]{lll}
(-+--)&=&-(V_{LR}+V_{LL})\mt\st \\
(-+-+)&=&+[E(V_{LR}+V_{LL})+k(V_{LL}-V_{LR})](1+\ct)\\
(-++-)&=&-[E(V_{LR}+V_{LL})-k(V_{LL}-V_{LR})](1-\ct)\\
(-+++)&=&+(V_{LR}+V_{LL})\mt\st\\
(+---)&=&-(V_{RL}+V_{RR})\mt\st\\
(+--+)&=&-[E(V_{RL}+V_{RR})-k(V_{RR}-V_{RL})](1-\ct)\\
(+-+-)&=&+[E(V_{RL}+V_{RR})+k(V_{RR}-V_{RL})](1+\ct)\\
(+-++)&=&+(V_{RL}+V_{RR})\mt\st.
\eaa
\label{vv_amp}
\eeq
Since the $\gamma$ and $Z$ exchange leads to helicity amplitudes of
the same form as those above, therefore we will use them to descibe
SM, vertex and vector four-Fermi operator effects.
Adopting for a notation $S\equiv S_{RR}$ and $T\equiv T_{RR}$
we obtain the following contributions to the differential cross sections
from scalar-tensor and vector operators, respectively:
\bea
\frac{d\sigma^{ST}}{d\ct}&=&\frac{N_C \bt}{32 \pi}\left[8|T|^2k^2\cts-
8\re(ST^\star)Ek\ct+|S|^2(2E^2-\mts)+4|T|^2\mts\right],\non \\
\frac{d\sigma^{VV}}{d\ct}&=&\frac{N_C \bt s}{256 \pi}\left[\left(
|A_L|^2+|A_R|^2+|B_L|^2+|B_R|^2\right)\bts\cts+\right.\\
&&\left.4\left(\re(A_L B_L^\star)-\re(A_R B_R^\star)\right)\bt\ct+\right.\non
\\
&&\left.2\left(|A_L|^2+|A_R|^2\right)
-\bts\left(|A_L|^2+|A_R|^2-|B_L|^2-|B_R|^2\right)\right],\non
\label{dsig}
\eea
where $N_C$ is a number of colours and $\bt=\frac{2}{\sqrt{s}}k$.
Above we used the following, more convenient notation:
\beq
\baa[c]{ll}
A_L=V_{LL}+V_{LR}&A_R=V_{RL}+V_{RR}\\
B_L=V_{LL}-V_{LR}&B_R=V_{RL}-V_{RR}.
\eaa
\eeq
One should notice that besides the SM $Z$ and $\gamma$ exchange
there are two sorts of contributions to $A_{L,R}$ and $B_{L,R}$:
these which enter as vertex corrections to the
$\ttbar Z$ and $\ttbar \gamma$ couplings (denoted as $A_{L,R}^{\gamma,Z}$
and $B_{L,R}^{\gamma,Z}$)
and those which emerge directly from vector type four-Fermi operators
(denoted by $A_{L,R}^{V}$ and $B_{L,R}^{V}$). The former are generated by
operators (\ref{ver_oper}) once the latter ones by operators
(\ref{fermi_oper}). {\it It should be noticed that $A_{L,R}^{\gamma,Z}$ and
$B_{L,R}^{\gamma,Z}$ are suppressed by the Z and/or $\gamma$ propagators
whereas
$A_{L,R}^{V}$ and $B_{L,R}^{V}$ are suppressed by the constant scale of new
physics $\lam$. This is why we should expect that at sufficiently high energy
vector type corrections shell dominate over vertex ones.}

Since an approximation $m_b=0$ have been adopted, there is no interference
between scalar-tensor and vector operators. However the SM contributions to
the amplitude do interfere with vector-type four-Fermi operators and those
which modify $\ttbar Z$ and $\ttbar \gamma$ vertices.
Although scalar-tensor operators do not interfere with the SM contributions
their effects are very relevant for sufficiently large CM energy
since the coefficients $S_{ij}$, $T_{ij}$ could be even of the order
of $\frac{1}{\lams}$.

\section{Experimental and Theoretical Constraints}

Coefficients of some of the relevant operators are restricted either by
experimental constraints or by theoretical requirements of naturality.
$\alpha^{(3)}_{\phi l}$,
coefficient of the operator $\ocal^{(3)}_{\phi l}$ (which enters our
calculations through corrections to the Fermi constant) is already restricted
by LEP data~\cite{lep}, which at the $ 3 \sigma $ level imply the following
constraints~\cite{gw}:
\beq
\ltv \gesim { 2.5 \sqrt{ \left| \alpha^{(3)}_{\phi l}\right| } }
\label{lepbounds}
\eeq
where $ \ltv $ is the scale of new physics in $\tev$ units.

Similarly the contributions to the oblique parameter
{\bf T}~\cite{peskin}
arise form $\ocal^{(3)}_\phi $, explicitly
\beq
\left| \delta {\bf T}  \right| =
{ 4 \pi \over \sw^2 } \left| \alpha^{(3)}_\phi
\right|
{ v^2 \over \Lambda ^2 } \lesim 0.4
\label{tbounds}
\eeq
This bound~\cite{pdb} implies $ \ltv \gesim 1.7 \sqrt{
\left| \alpha^{(3)}_\phi \right| } $ (at $ 3 \sigma $).

However, $\alpha^{(3)}_{ll}$ and $\alpha^{(1)}_\phi$~\footnote{None of
the high precision measurements constrain $ \alpha^{(1)}_\phi $ since,
without direct observation of the Higgs, the tree-level effects of this
operator are absorbed in the wave function renormalization of the scalar
doublet.} both contributing
to corrections to our input parameters are experimentally unconstrained.

It is easy to notice that scalar and tensor type four-Fermi operators emerging
after applying Fiertz transformation to $\ocal_{lq}$ and $\ocal_{lq'}$ would
contribute at the one-loop level to the electron mass and anomalous magnetic
moment of the electron:
\beq
\delta m_e^S \sim \mt S \frac{\mts}{\lams} \lsp
\delta a^T_e \sim T \frac{m_e\mt}{\lams}.
\label{naturality}
\eeq
Although, one can imagine some mechanisms to cancel above contributions,
preserving non-zero $S$ and $T$,
however here, presenting numerical results we will assume $S=T=0$ to
avoid any fine-tuning necessary to overcome the above constraints.

\section{Results and Perspectives}

\begin{figure}
\vspace*{13pt}
\vspace*{3.2truein}
%ORIGINAL SIZE=1.6TRUEIN x 100% - 0.2TRUEIN
\includegraphics{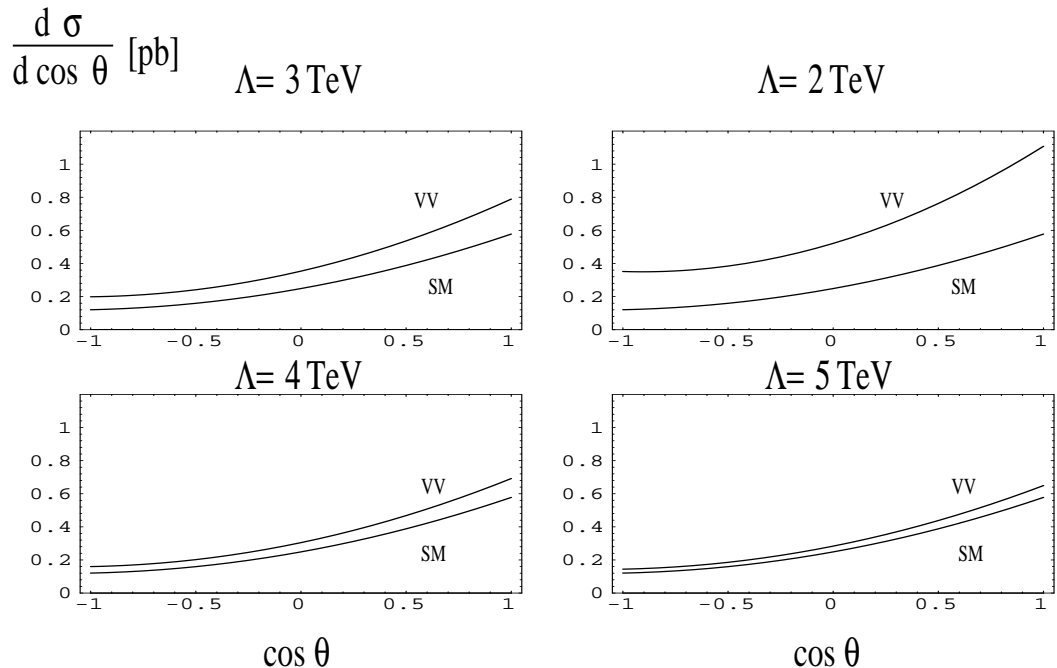}
\caption{The differential cross section $\dsig$ as a function of $\ct$ for
$s =.5^2\tev^2 $ for $\lam=2,3,4,5\tev$. $VV$ and $SM$ denote the prediction
with and without vector-type four-Fermi operators, respectively.}
\label{fig6}
\end{figure}

We focus here on two observables, the total cross section $\sig$
and $\fba$ for $\eett$. However, it is instructive first to look at
a differential cross section $\dsig$ presented in the Fig.~\ref{fig6}.
Since we would like to emphasize the effects of four-Fermi operators,
in numerical calculations we assumed always $S=T=A_{L,R}^{\gamma,Z}=
B_{L,R}^{\gamma,Z}=0$ and
$A^V_{L,R}=B^V_{L,R}=-1/\lams$,
also corrections to $G_F$, $m_Z$ and $\alpha_{QED}$ are suppressed for this
purpose. Although corrections to the $\dsig$ can be substantial (even $100\%$
for $\lam=2\tev$), however it is seen that the main effect consist of rescaling
the angular distribution.

\begin{figure}
\vspace*{13pt}
\vspace*{3.2truein}
%ORIGINAL SIZE=1.6TRUEIN x 100% - 0.2TRUEIN
\includegraphics{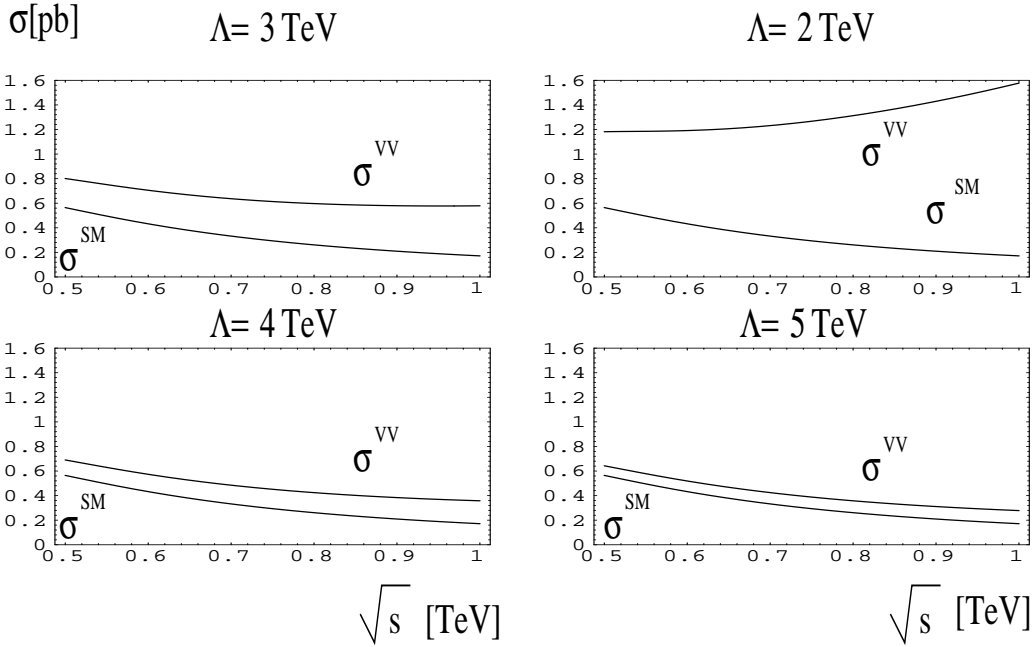}
\caption{The total cros section $\sig$ as a function of the CM energy
for $\lam=2,3,4,5\tev$. $VV$ and $SM$ denote the prediction
with and without vector-type four-Fermi operators, respectively.}
\label{fig5}
\end{figure}

The total cross section $\sig$ as a function of the
center of mass energy is shown in Fig.~\ref{fig5}. As we have already
anticipated the corrections are rising with energy what is an effect of
relative enhancement of four-Fermi interactions.

There are two quantities relevant for experimental potential of NLC,
namely the total integrated luminosity $L$ and the tagging efficiency
for an observation
of $\ttbar$ pairs $\varepsilon_{tt}$. Since they both enter the statistical
significance in a combination $\sqrt{\varepsilon_{tt} L}$ it will useful to
adopt
a notation $\epsl\equiv\sqrt{\varepsilon_{tt} L}$ and parameterize our results
in terms of $\epsl$.
Below we present a table showing $\varepsilon_{tt}$ and $L$ corresponding to
$\epsl=30,50,100\pbarn^{-1/2}$.
\begin{center}
\begin{tabular}{||l||r|r|r|r|r||}
\hline \hline
$L[10^4\pbarn^{-1}]$&1.&2.&3.&4.&5.\\ \hline
$\varepsilon_{tt}$(for $\epsl=30\pbarn^{-1/2}) $&.09&.05&.03&.02&.02\\ \hline
$\varepsilon_{tt}$(for $\epsl=50\pbarn^{-1/2}) $&.25&.13&.08&.06&.02\\ \hline
$\varepsilon_{tt}$(for $\epsl=100\pbarn^{-1/2})$&1.0&.50&.33&.25&.20\\
\hline \hline
\end{tabular}
\end{center}
Since $L=2\times10^4\pbarn^{-1}$ looks presently as realistic yearly available
luminosity, we can see from the above table that $\epsl=50\pbarn^{-1/2}$
would require only $13\%$ for $\ttbar$ tagging efficiency $\varepsilon_{tt}$.
Some rough estimations of the efficiency are available in the
literature~\cite{efficiency} for the most frequent final state topologies:
6 jets ($BR \sim 50\%$) and 4 jets + 1 charged lepton ($BR \sim 40\%$).
For the 6 jet channel
$\varepsilon_{tt}$ is expected to be around $30\%$,
whereas for 4 jet + 1 lepton $15\%$ should
be obtained. Therefore we can conclude that already $\epsl=50\pbarn^{-1/2}$
can serve as a realistic estimation of the detection potential of the NLC.

The quantity which provides the relevant for us information is the statistical
significance of the non-standard physics effects in $\eett$. For the total
cross section $\sig$ statistical significance is given by the following
formula:
\beq
\nsdsig=\epsl\frac{\sig-\sig^{SM}}{\sqrt{\sig}},
\label{nsd_sig}
\eeq
where $\sig\equiv\sigma^{ST}+\sigma^{VV}$.
For the forward backward asymmetry we obtain:
\beq
\nsdasy=\frac{\fba-\fba^{SM}}{\sqrt{1-\fba^2}}\sqrt{\sig}\;\epsl.
\label{nsd_bfa}
\eeq

We must remember that we are not
allowed to trust our lowest order effective Lagrangian calculation whenever
relative corrections are greater than, say $10\%$, therefore it is useful
to define
\beq
\kappa\equiv\frac{\sig-\sig^{SM}}{\sig^{SM}}\;100\%\lsp
\eta\equiv\frac{\fba-\fba^{SM}}{\fba^{SM}}\;100\%.
\label{corrections}
\eeq

\begin{figure}
\vspace*{13pt}
\vspace*{3.2truein}
%ORIGINAL SIZE=1.6TRUEIN x 100% - 0.2TRUEIN
\includegraphics{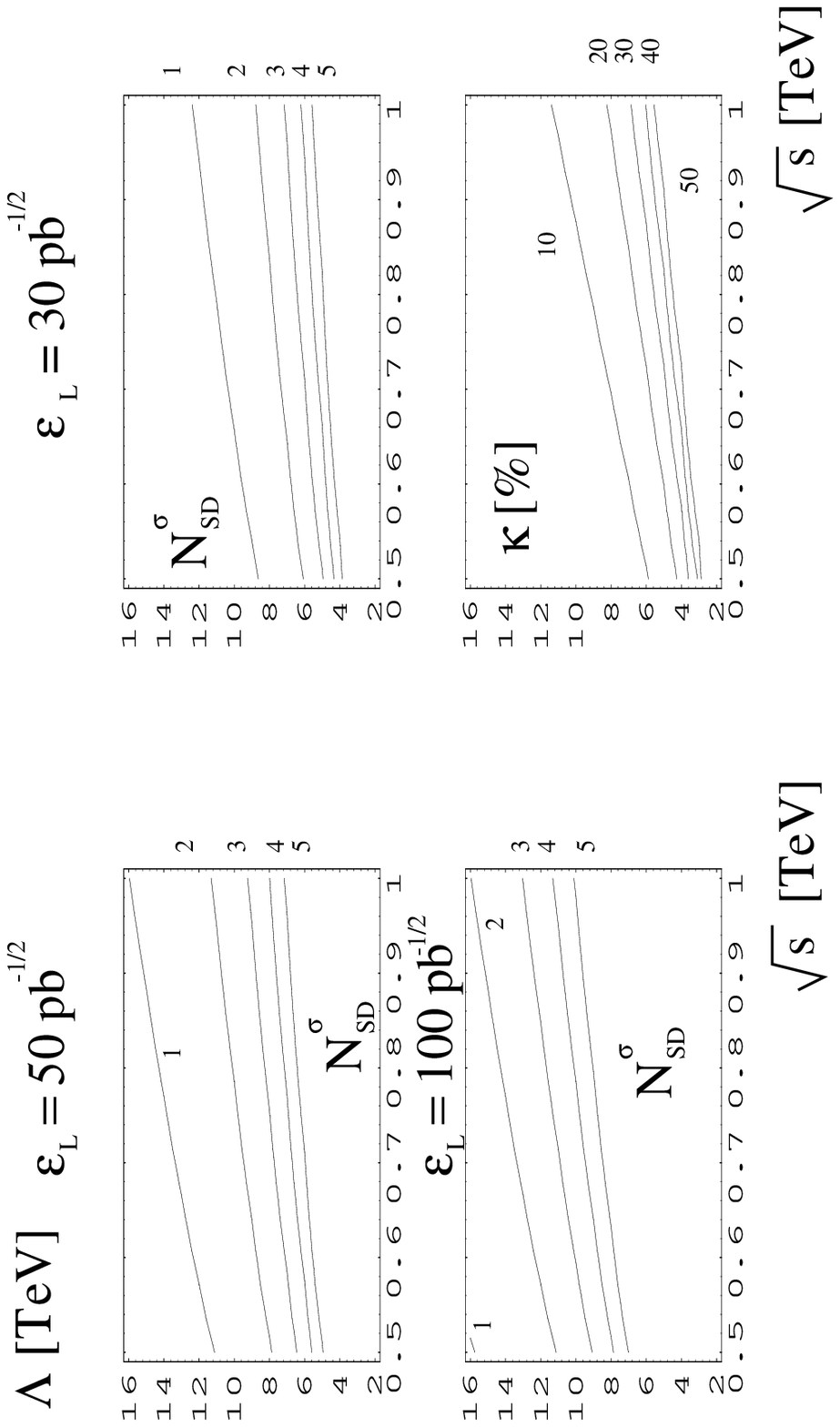}
\caption{Contour plots for the statistical significance $\nsdsig$ for an
observation of non-standard effects in the total cross section for
$\epsl=30,50,100 \pbarn^{-1/2}$.
Contour plot for the relative correction $\kappa$ to $\sig$ is shown
in the lower righ plot.}
\label{fig1}
\end{figure}

In the Fig.~\ref{fig1} we present for $\epsl=30,50,100\pbarn^{-1/2}$
contour plots for
the $\nsdsig$ and also for $\kappa$ in the $\lam$-$\sqrt{s}$ plane.
Looking at $\nsdsig$ plots it is instructive to check whether for
considered $\lam$ and $\sqrt{s}$ we are still in the perturbative region.
It can be seen that for $\epsl=100\pbarn^{-1/2}$ the non-standard effects can
be observed even at the $5\sigma$ level in the entire plane keeping
$\kappa$ below $10\%$. For $\epsl=50\pbarn^{-1/2}$ and
$\sqrt{s}< .7\tev$ effects could be observable at $3\sigma$ level, for
larger $\sqrt{s}$ corrections are greater then $10\%$.
For $\epsl=30\pbarn^{-1/2}$ only $1\sigma$ would be available for the entire
range of $\sqrt{s}$. These plots are designed to answer the question
{\it what are the machine parameters necessary to test the nonstandard effects
at a desired confidence level}.

\begin{figure}
\vspace*{13pt}
\vspace*{3.2truein}
%ORIGINAL SIZE=1.6TRUEIN x 100% - 0.2TRUEIN
\includegraphics{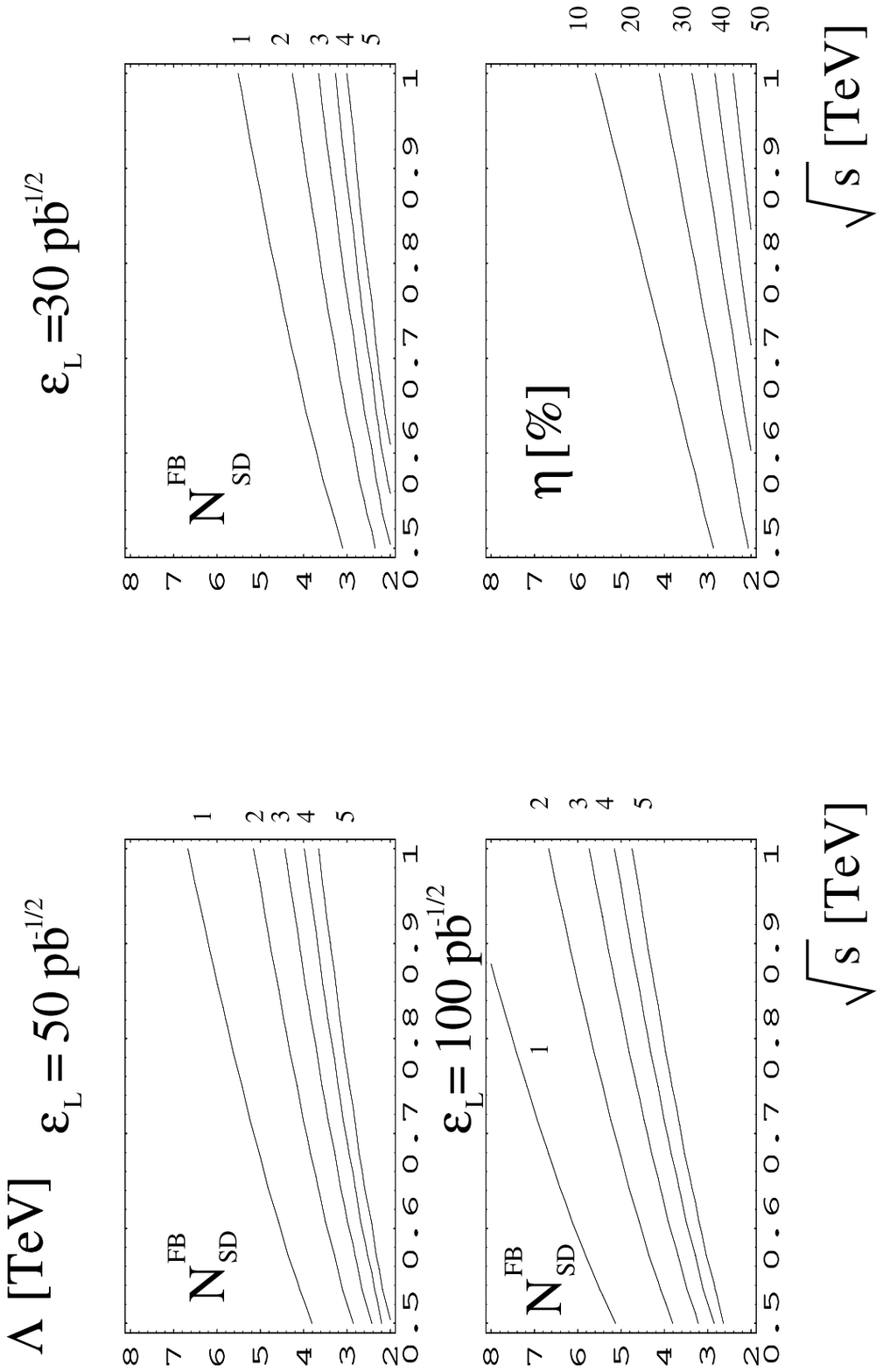}
\caption{Contour plots for the statistical significance $\nsdasy$ for an
observation of non-standard effects in the forward-backward asymmetry
$\epsl=30,50,100 \pbarn^{-1/2}$.
Contour plot for the relative correction $\eta$ to $\fba$ is shown
in the lower righ plot.}
\label{fig2}
\end{figure}

The same attitude is assumed the Fig.~\ref{fig2}, where we plot the
statistical significance $\nsdasy$ for the forward-backward asymmetry $\fba$.
We can see from the plots that here it is much less promising to
observe an effect
of non-standard physics. Even having $\epsl=100\pbarn^{-1/2}$ one can only
expect effects up to $3\sigma$ level, staying below $\eta=10\%$. It is easy to
understand the reason for that, as we have already noticed that the four-Fermi
operator effect the $\dsig$ mainly by rescaling it, therefore the $\fba$
is not corrected that much as the total cross section $\sig$ itself.

If one decided to keep $S$ and $T$ non-zero, than it could look attractive to
consider polarized $\epem$ beams in order to suppress the SM contribution
relative to non-standard scalar and/or tensor operator effects. However,
as it has been checked by a direct calculation,
with the same polarization for
both $e^+$ and $e^-$ beams
this strategy is not very promising since the relative
corrections to the $\sig$ and $\fba$ grows faster with the polarization
then the corresponding
statistical significances and therefore we easy enter a region in
the $\lam$-$\sqrt{s}$ plane where even for $\epsl=100\pbarn^{-1/2}$
tests at the
level of $5\;\;\sigma$ would correspond to relative corrections close to
$20\;\%$ which is already too large to trust our effective-Lagrangian
tree-level computations.

%\begin{figure}[t]  % t means top of the page, b- bottom, (..?)
%\postscript{fig1_g.ps}{0.65}              %**epsf**
%\vspace{-0.25cm}
%\caption{ }
%\label{fig1}
%\end{figure}

\section{Conclusions}
We emphasize here that a consistent analysis of the process $\eett$
can not be restricted to non-standard vertex corrections as it is
often done in the existing literature.
It has been shown that even considering only vector-type four-Fermi operators
the prediction for the total cross section can be substantially modified
allowing for test of non-standard physics of a scale of about $5\tev$.
Ignoring theoretical prejudices and keeping $S$ and $T$ non-zero
beyond the \sm\ effects are even more pronounced.
It has been checked that looking for non-standard physics in
$\sig$ and $\fba$ it is more convenient to have a machine with larger
$\epsl\equiv\sqrt{\varepsilon_{tt}L}$ than one with polarized beams.

\vspace{1cm}
\centerline{\bf Acknowledgments}
\vspace{.5cm}

We thank R. Vega for his interest at the beginning of this research and
J. Wudka for useful comments.
We thank the HEP theory group at Argonne National Laboratory and
and the theory group of
Max-Planck-Institute f\"ur Physik und Astrophysik for a hospitality and
support during the course of this research.


\begin{thebibliography}{99}

\bibitem{CDF}
F. Abe \etal, CDF Collaboration, \prd{50}{1994}{2966};\\
F. Abe \etal, CDF Collaboration, Fermilab-Pub-95/022-E;\\
S. Abachi \etal, D0 Collaboration, Fermilab-Pub-95/028-E.

\bibitem{top}
W. Bernreuther nad O. Nachtmann, \plb{268}{1991}{424};\\
D. Atwood and A. Soni, \prd{45}{1992}{2405};\\
G.L. Kane, G.A. Ladinsky and C.-P. Yuan, \prd{45}{1992}{124};\\
C.-P. Yuan, \prd{45}{1992}{782};\\
W. Bernreuther \etal, \plb{279}{1992}{389};\\
G.A. Ladinsky and C.-P. Yaun, \prd{49}{1994}{4415};\\
F. Cuypers and S.D. Rindani, \plb{343}{1995}{333};\\
C.-P. Yuan, \mpla{10}{1995}{627}.

\bibitem{leffref}
J.M. Cornwall \etal, \prd{10}{1974}{1145};\\
M. Chanowitz and M.K. Gaillard, \npb{261}{1985}{379};\\
H. Georgi, \npb{361}{1991}{339}, \npb{363}{1991}{301};\\
J. Wudka, \ijmpa{9}{1994}{2301}.

\bibitem{gi}
M.B. Einhorn and J. Wudka, in {\it Workshop on Electroweak Symmetry
  Breaking}, Hiroshima, Nov. 12-15 (1991); in {\it Yale Workshop on Future
  Colliders,} Oct. 2-3 (1992);\\
M.B. Einhorn, in {\it Workshop on Physics and Experimentation with
  Linear $e^+e^-$ Colliders},  Waikoloa, Hawaii, April 26-30, 1993;\\
J. Wudka, in {\it Electroweak Interactions and Unified Theories,}
  XXVIII Recontres de Moriond Les Arcs, Savoie, France, March 13-20 (1993).

\bibitem{Appelquist} T. Appelquist and J. Carazzone, \prd{11}{1975}{2856}.

\bibitem{bw} W. B\"uchmuller and D. Wyler, \npb{268}{1986}{621};\\
C.J.C.  Burges and H.J. Schnitzer, \npb{228}{1983}{464};\\
C.N. Leung \etal, \zfp{31}{1986}{433}.

\bibitem{Arzt.et.al.} C. Arzt \etal, \npb{433}{1995}{41}.

\bibitem{others}
A. de R\'ujula \etal, \npb{384}{1992}{3};\\
C. Grosse-Knetter \etal, \zfp{60}{1993}{375};\\
K. Hagiwara \etal, \plb{318}{1993}{155};\\
M. Bilenkii \etal, \npb{419}{1994}{240};\\
G. Gounaris \etal, \plb{338}{1994}{51}.

\bibitem{lep} The LEP collaborations, report CERN/PPE/94-187.

\bibitem{gw} B. Grz\c{a}dkowski and J. Wudka, preprint IFT-2/95, UCR-T133,
hep-ph/9502415, to appear in {\it Phys. Lett.} {\bf B}.

\bibitem{peskin} M. Peskin and T. Takeuchi, \prl{65}{964}{1990}.

\bibitem{pdb} Particle Data Book, \prd{50}{1173}{1994}.

\bibitem{efficiency} K. Fujii, in {\it Workshop on Physics and
Experiments with Linear Colliders}, Saariselka, Finland, September 9-14,
1991;\\
 P. Igo-Kemens, in {\it Workshop on Physics and
Experimentation with Linear $e^+e^-$ Colliders},  Waikoloa,
Hawaii, April 26-30, 1993.

\end{thebibliography}
\end{document}